\documentclass[preprint2]{aastex}

\shorttitle{Solar-cycle variation of sound speed}
\shortauthors{Rabello-Soares}

\begin{document}

\title{Solar-cycle variation of sound speed near the solar surface}

\author{M. C. Rabello-Soares}
\affil{W. W. Hansen Experimental Physics Laboratory, Stanford University, Stanford, CA 94305, USA}
\email{cristina@sun.stanford.edu}

\begin{abstract}

We present evidence that the sound-speed variation 
with solar activity
has a two-layer configuration, 
similar to the one observed below an active region,
which consists of 
a negative layer near the solar surface and
a positive one in the layer 
immediately below the first one.
Frequency differences 
between the activity minimum and maximum of solar cycle 23,
obtained applying global helioseismology 
to the Michelson Doppler Imager (MDI) on board SOHO,
is used to determine the sound-speed variation 
from below the base of the convection zone to 
a few Mm below the solar surface.
We find that 
the sound speed at solar maximum is smaller than at solar minimum
at the limit of our determination (5.5 Mm).
The min-to-max difference decreases in absolute values until $\sim7$ Mm.
At larger depths,
the sound speed at solar maximum is larger than at solar minimum
and their difference increases with depth until $\sim10$ Mm.
At this depth, the relative difference ($\delta c^2 / c^2$) is less than half of the value observed at the lowest depth determination. 
At deeper layers,
it slowly decreases with depth until there is no difference between maximum and minimum activity.

\end{abstract}


\section{Introduction}

The Sun has an approximately 22-year magnetic cycle, where the dipolar magnetic field at the solar poles reverses each 11 years or so. During the reversal, the magnetic activity is at a minimum and very few sunspots are visible on the Sun, sometimes none can be seen. At the maximum in the cycle, the number of sunspots visible on the Sun can be more than 100 at one time. 
A typical sunspot has a lifetime of a few weeks.
The radio, ultraviolet and X-ray emission 
of the Sun also increase significantly during solar maximum.
The total solar irradiance too has a strong correlation with the sunspot number.
Significant flares 
occur at all phases of the sunspot cycle, but
the number of M-class and X-class flares 
tends to follow the sunspot number.
There are several proxies of solar activity currently used, probing the solar
atmosphere at different heights \citep[see, for example,][]{jain09,hathaway2010}.
The most accepted scenario is that the solar cycle is 
produced by dynamo processes within the Sun and it is
driven by differential rotation and convection.

As initially reported by \cite{woodard85},
it is now well established that
the frequency of the solar acoustic modes is strong correlated
with solar activity.
This has been observed for modes with
degree up to 3
\citep[e.g.,][]{salabert04, chaplin2007},
up to about 200
\citep[e.g.,][]{bachmann93, jain03, antia03, dziembowski05, jain09}
and up to 1000 \citep[e.g.,][]{rhodes02, rabello06, rabello11, rhodes11}.
The observed mode frequencies can be inverted to obtain the sound speed 
in the solar interior where the waves travel through.
However, in analyzing the changes in mode frequency with the magnetic solar cycle,
we have to keep in mind that
magnetic fields can change not only the thermodynamic structure of the medium the waves travel through, which, in turn, changes the frequencies of the waves
but also
the plasma waves can be directly affected by the magnetic fields through the Lorentz force.
In short, the modification to the acoustic mode frequencies results from both
structural and non-structural effects of the magnetic fields \citep[see][]{Lin06}.

The measured solar-cycle frequency shifts depend strongly only on the frequency of the mode 
after they have been weighted by the mode inertia \citep{libb90}.
This indicates that the dominant structural changes during the solar cycle, 
so far as
they affect the mode frequencies, occur near the surface.
At the outermost layers, the timescale of the heat transfer is smaller than the period of the oscillation and they are not adiabatic anymore.
At the moment, there is no general agreement as to the
precise physical mechanism near the surface that gives rise to the frequency variation.
\citep[see, for example,][]{li03}.
Several authors confirmed that all or most of the physical changes were
confined to the shallow layers of the Sun
\citep[e.g.,][etc]{basu02, darwich02, dziembowski05, rabello09}.
However, \cite{baldner08} 
observed a small, but significant change in the sound speed at the base of the convective zone and for $r \gtrsim 0.9$ $R_\odot$.
A change in the interior structure was also found by \cite{basumandel04},
specifically
a variation in adiabatic index, $\Gamma_1$, 
near the second helium ionization zone
($\sim0.98$ $R_\odot$).
On the other hand, changes in zonal and meridional flows
correlated with solar activity have been observed by several authors
\citep[][etc]{schou99, howe00, vorontsov02, howe05, basu06}.

Local helioseismology techniques have shown that 
below an active region there is a
two-layer structure with a negative variation of the sound speed
in a shallow subsurface layer and a positive variation in the deeper interior
relative to a quiet Sun region \citep[see][and references within]{kosovichev00,basu04,bogart08,kosovichev11}.
A similar variation has been observed for the adiabatic index, $\Gamma_1$
below an active region \citep{basu04, bogart08}.
Because
global activity levels are much smaller than in active regions,
a similar 
structure variation
would be much smaller in a global scale, if it occurs at all.

\cite{rabello11} 
estimated the frequency differences between the activity minimum and
maximum 
by carefully fitting the variation of the 
acoustic mode frequency with solar activity
for modes with $20 \le l \le 900$ observed by MDI/SOHO.
Here, these 
frequency differences (described in Section 2) are inverted to 
look for the interior sound-speed variation with solar cycle.

\section{Data used}

Full-disk Doppler images obtained at a one-minute cadence by
MDI Dynamics and Structure observing modes were analyzed \citep{scherrer95}.
The first one has higher spatial resolution ($1024^2$ pixels) and
is available every year for two or three months of continuous data.
Data for the years 1999 to 2008 were used. 
The length of the time series varies from
38 days (in 2003) to 90 days (in 2001).
The second observing mode is available year around, but is subsampled in order to fit the limited telemetry (modes up to $l \approx 250$). It is divided in 72 day time series and data from early 1996 to April 2010 were used.
They will be called the Dynamics and Structure sets from now on.
Each velocity image is
decomposed into spherical-harmonic components \citep{schou02}.

Two different peak-fitting algorithms were used to fit the power spectra and obtain the mode frequencies. 
One of them known as the MDI peak-fitting method is described in detail 
by \cite{schou92} \citep[see also][]{schou02,larson08,larson09}.
A spherical-harmonic decomposition is not orthonormal over 
the solar surface that can be observed from a single view point 
resulting in what is referred as spatial leakage.
At high degrees, the spatial leaks lie closer in frequency 
(due to a smaller mode separation) 
resulting in the overlap of the target mode with the spatial leaks that merges individual peaks into ridges, making it more difficult to estimate unbiased mode frequencies.
The MDI peak-fitting method works very well for low and medium-$l$ modes where the leaks are well separated from the target mode. 
For high-$l$ modes, which have spatial leaks overlapping the target mode and blended in a single ridge, 
\cite{korzennik98} method was used \citep[for details, see][]{rabello08b}.
The fitting was carried out only for every tenth $l$ 
for modes with $100 \le l \le 900$.
This method was applied only to the Dynamics time series which were Fourier transformed in small segments (4096 minutes) whose spectra were averaged to produce an averaged power
spectrum with a low but adequate frequency resolution to fit the ridge while reducing the realization noise. 
These two peak-fitting methods will be called ML and HL respectively.

\begin{figure}[h]
\includegraphics
[scale=0.4]
{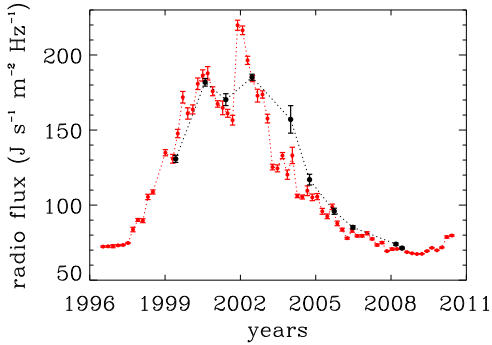}
\caption{\label{radioflux} Mean radio flux for the observational periods in the Dynamics set (in black) and in the Structure set (in red).
The error bars correspond to the error of the mean. 
}
\end{figure}

The ML method was applied to both observing sets and the HL only to the Dynamics set.
In summary, three types of data were used in this work: 
(1) frequencies obtained by both ML and HL methods applied to the Dynamics set (``Dynamics ML+HL''); (2) frequencies obtained by only the ML method applied to the Dynamics set (``Dynamics ML''); and (3) frequencies obtained by only the ML method applied to the Structure set (``Structure ML'').

The estimated mode frequencies, $\nu_{nlm}$, were parameterized in terms of Clebsch-Gordan coefficients \citep{ritz91}. The central frequency, i.e., the frequency free of splitting effects, is taken to be the frequency given by $m=0$ in the parameterization.
For each of these three sets,
the central frequency, $\nu_{nl}$, for each time period was fitted assuming a
linear and a quadratic relationship with the correspondent solar-activity index and using a weighted least-squares minimization. For a detailed description, see \cite{rabello11}. 
The solar radio 10.7 cm daily flux (NGDC/NOAA) was used as the solar-activity proxy (Figure~\ref{radioflux}).

Although, 
there is a very high linear correlation of the mode frequency variation with several solar-activity indices,
deviations from a simple linear relation have been reported \citep[see, for example,][]{chaplin2007}. More recently,
\cite{rabello11} found some evidence of a quadratic relationship
indicating a saturation at high solar activity ($\sim350 \times 10^{-22}$ J s$^{-1}$ m$^{-2}$ Hz$^{-1}$).
A similar effect has been
seen in frequencies at activity regions with a large surface magnetic field using ring analysis
(Basu et al. 2004).
The frequency shifts obtained using the quadratic fitting were used here.

\begin{figure}[h]
\includegraphics
[scale=0.4]
{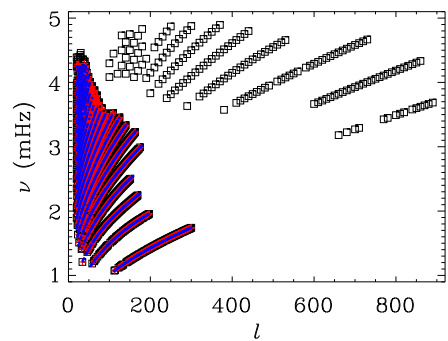}
\caption{\label{lnuinv}
Modes used in the inversion for the 
Dynamics ML+HL, Dynamics ML and Structure ML
sets are shown as black squares, red crosses and
blue circles, respectively.
} 
\end{figure}

Modes with a negative linear coefficient in the quadratic fitting were
excluded from the present analysis.
They also have a positive quadratic coefficient, indicating that they are less
affected by low and medium levels of solar activity.
They correspond to 5\%, 11\% and 16\% of the modes of the Structure ML, Dynamics ML and Dynamics ML+HL sets, respectively.
In the Dynamics ML+HL mode set, the excluded modes make for the lack of modes in the middle of the $l-\nu$ diagram in Figure~\ref{lnuinv}.
The $f$-modes obtained with the HL method (about 20 modes)
have a distinctive behavior from those obtained with the ML method \citep[see Figure~7 in][]{rabello11}.
It is not clear if this behavior is an artifact of the frequency determination or not. They were excluded from our analysis.

The solar activity minimum-to-maximum frequency shift $\delta\nu^{sc}_{nl}$ 
is defined as the difference between the fitted frequency at the maximum and at the minimum of solar activity.
The minimum and maximum activity 
correspond to the Dynamics observing periods centered around 
April 2008 and May 2002 respectively (Figure~\ref{radioflux}). 
These observing periods were also used to calculate the frequency difference for the Structure set. 
The corresponding radio fluxes are 71 and 185 J s$^{-1}$ m$^{-2}$ Hz$^{-1}$ for both sets.
The minimum-to-maximum frequency shifts $\delta\nu^{sc}_{nl}$ 
obtained using a linear or a quadratic fit to the solar activity are very similar.

\section{The surface term}

The scaled frequency differences $Q_{nl} \, \delta\nu_{nl} / \nu_{nl}$ 
(between two models or between the Sun and a model)
can be expressed as 
$H_1(\nu_{nl}/L) + H_2(\nu_{nl})$, where 
$H_1$ is the contribution of the interior sound-speed difference, 
$H_2$ of the differences in the surface layers,
$Q_{nl}$ is the mode inertia normalized by the inertia of a radial mode of the same frequency and $L = \sqrt{l(l+1)}$ \citep{jcd1989}. 
$H_2$ is also called the surface term, $F_{\rm{surf}}$.

\begin{figure}[h]
\includegraphics
[scale=0.4]
{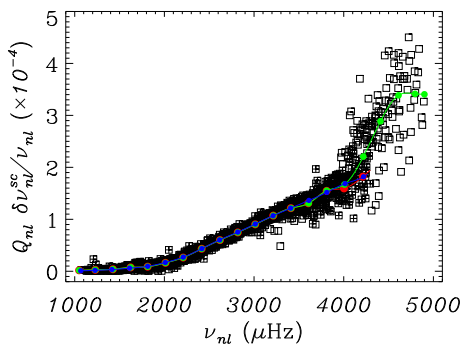}
\caption
{
Scaled minimum-to-maximum frequency differences obtained using Dynamics ML+HL (black squares), Dynamics ML (black crosses) and Structure ML (small black circles) sets.
Green, red and blue circles correspond to the averaged values for Dynamics ML+HL, Dynamics ML and Structure ML, respectively.
The solid lines are the interpolated splines representing the surface term, $F_{\rm{surf}}$.
}
\label{fig:fitting}
\end{figure}

\begin{figure}[h]
\includegraphics
[scale=0.4]
{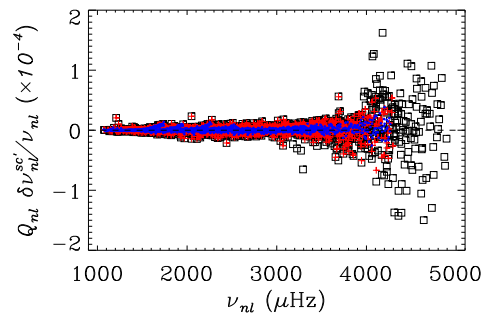}
\caption
{Residuals, $Q_{nl} \, \delta\nu^{sc'}_{nl} / \nu_{nl}$, obtained after subtracting $F_{\rm{surf}}$.
The black squares, red crosses and blue circles correspond to Dynamics ML+HL, Dynamics ML and Structure ML, respectively.
}
\label{fig:residuals}
\end{figure}

Figure~\ref{fig:fitting} shows the scaled frequency differences between solar minimum and maximum,  $Q_{nl} \, \delta\nu^{sc}_{nl} / \nu_{nl}$,
obtained using Dynamics ML+HL (black squares), Dynamics ML (black crosses) and Structure ML (small black circles) sets.
They are
basically a function of frequency.
The near-surface contribution can be removed
by subtracting a smooth function of frequency, $F_{\rm{surf}}(\nu)$, from the scaled frequency shifts,
thus
allowing the search for structure variations in the solar interior with the solar cycle.
$Q_{nl} \, \delta\nu^{sc}_{nl} / \nu_{nl}$ were averaged
over 200-$\mu$Hz intervals.
In each interval, outliers large than 3-$\sigma$ were removed and the weighted average calculated.
The results are represented in Figure~\ref{fig:fitting} as 
green, red and blue 
circles respectively for Dynamics ML+HL, Dynamics ML and Structure ML sets.
A cubic spline interpolation was applied to the 17 and 21 frequency-knots for the ML and ML+HL sets, respectively (solid lines in Figure~\ref{fig:fitting}).
Figure~\ref{fig:residuals} shows the residuals after subtracting
the surface term given by the splines.

\section{Results}

\begin{figure}[h]
\includegraphics
[scale=0.4]
{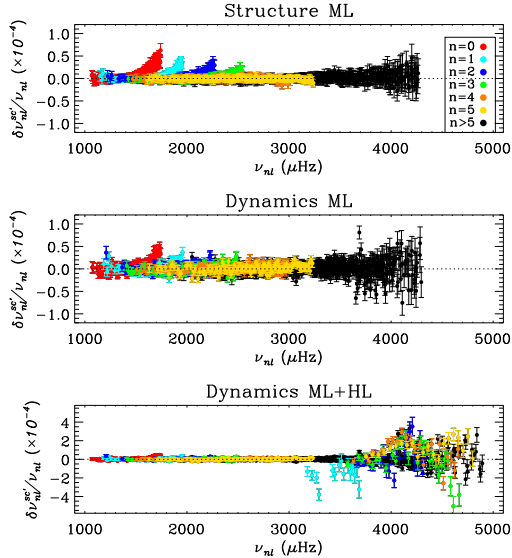}
\caption
{
Relative frequency differences between solar maximum and minimum, after removing the surface term, used in the sound-speed inversion.
}
\label{fig:delnu_n}
\end{figure}

A nonasymptotic inversion technique can be applied to infer the sound speed (and density) inside the Sun using the frequency differences between the Sun and a reference solar model \citep[e.g.][and references within]{antia94}.
Here the solar-cycle frequency shifts, $\delta\nu^{sc'}_{nl}$, were inverted to determine the differences in sound speed between minimum and maximum solar activity inside the Sun.
The inversion has been carried out using the 
multiplicative optimally localized averages technique 
\citep{gough85}. 
The relative frequency shifts $\delta\nu^{sc'}_{nl} / \nu_{nl}$ are plotted in Figure~\ref{fig:delnu_n}. They are small
indicating small changes in the solar structure with the solar cycle, if any at all.

\begin{figure}[h]
\includegraphics
[scale=0.4]
{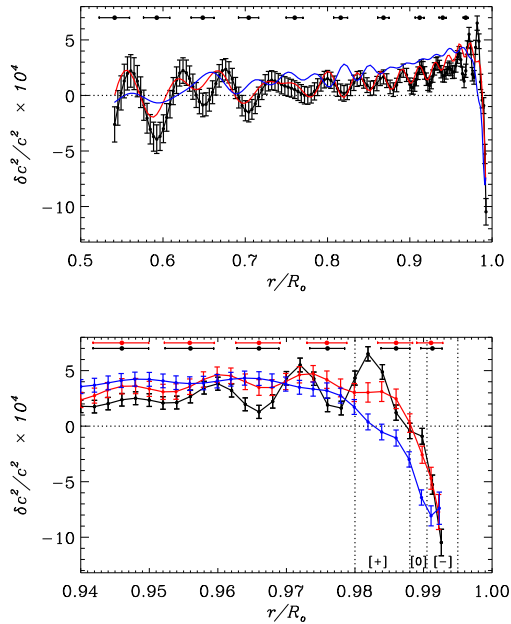}
\caption
{
Inferred sound-speed differences
obtained using Dynamics ML+HL (black circles), Dynamics ML (red line) and
Structure ML (blue line) sets.
The bottom panel shows the solution near the surface in more detail.
The vertical dotted lines mark approximately the regions that correspond
to positive, zero and negative sound-speed variations below an active region
in relation to a quiet region obtained by ring analysis \citep{bogart08}.
}
\label{fig:solution}
\end{figure}

The inferred sound-speed differences 
are shown in Figure~\ref{fig:solution}.
The results obtained using Dynamics ML+HL (black), Dynamics ML (red) and Structure ML (blue) sets
follow the same general trend. The sound-speed difference increases slowly with radius until about 0.98 $R_\odot$ after which it decreases sharply.
The sound-speed difference is larger than 6-$\sigma$ 
at its maximum for all three sets.
The inversion results are identical if a linear fit of the mode frequencies to the solar-activity index is used instead of a second-degree polynomial.

\begin{figure}[h]
\includegraphics
[scale=0.4]
{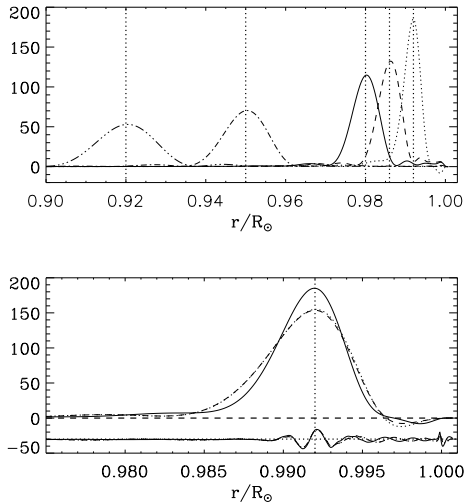}
\caption
{Top: Samples of averaging kernels at five different radii near the surface for the Dynamics ML+HL set.
The target radius is shown by a vertical dotted line.
Bottom: Averaging kernels at $r = 0.992$ $R_\odot$ for Dynamics ML+HL (solid line), Dynamics ML (dashed line) and Structure ML (dotted line) sets.
The cross term kernel 
is also plotted for all three sets. It is multiplied by 100 and displaced vertically by -30 for visualization purpose.
\label{avgker}
}
\end{figure}

The solution errors for the Dynamics ML+HL set are smaller by as much as 25\% than for the other two sets for $r > 0.85$ $R_\odot$
and 15\% for $r < 0.85$ $R_\odot$.
Also, the averaging kernels for Dynamics ML+HL 
are slightly more localized close to the surface than the other two sets
due to the inclusion of high-$l$ modes.
In Figure~\ref{fig:solution}, 
the averaging-kernel widths (defined as the difference between the first and third quartile points) are given by the horizontal error bars for a few selected locations. 
In the top panel, the horizontal bars are the same for all three sets. 
In the bottom panel, 
they are in black for Dynamics ML+HL and
in red for Dynamics ML and Structure ML sets, since they are very similar.
Nevertheless, the averaging kernels are well localized and have almost no side lobes for the inferred solution from 0.53 to 0.992 $R_{\odot}$ for all three sets 
(Figure~\ref{avgker}).
The cross-term kernel
measures the influence of the contribution from density on the inferred sound speed and it is insignificant in this range (bottom panel).
We did not include modes with $l < 20$ in our analysis since we
are not interested here in the solar deeper layers. 
For $r > 0.992$ $R_{\odot}$,
the averaging kernels become increasingly asymmetric and develop large side lobes.

To check that the surface term is properly removed and is not causing
the observed sound-speed variation, 
the surface term was estimated using four times more frequency-knots than used in Figure~\ref{fig:fitting} (i.e., using 50 $\mu$Hz intervals).
In this case, $F_{\rm{surf}}$ is not a smooth function of frequency anymore.
The inversion of the residuals is still similar to Figure~\ref{fig:solution}. 
The maximum sound-speed variation around 0.98 $R_\odot$ is 
still statistically significant, it is  
larger than 4-$\sigma$ (larger than 6-$\sigma$ for the Dynamics ML+HL set).

\begin{figure}[h]
\includegraphics
[scale=0.4]
{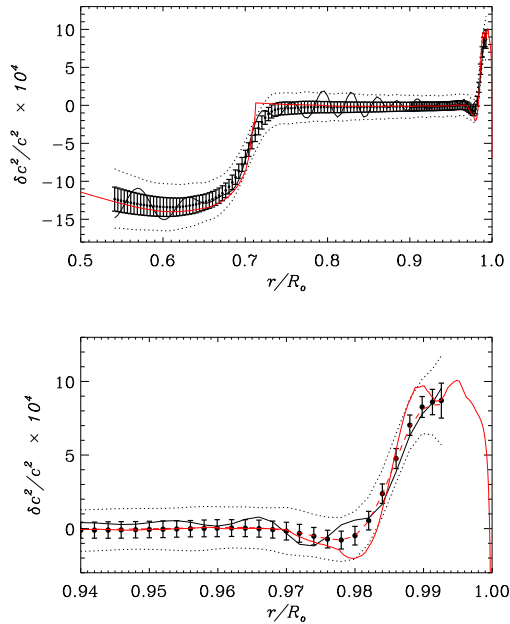}
\caption
{
Sound-speed differences
between two solar models 
through most of the solar interior (top) and, 
in detail, near the surface (bottom).
The solid red line is the true difference between the models,
while the dashed red line (bottom panel) is the difference weighted by the averaging kernels.
The black circles represent the inversion results 
using the observed mode set for Dynamics ML+HL
and the solid black line 
the inversion results with random noise added to the theoretical frequencies.
The two dotted black lines correspond to the average of 1000 realizations plus and minus 2.5-$\sigma$.
\label{artificial}
}
\end{figure}

Figure~\ref{artificial} shows the sound-speed difference between two different solar models (solid red line) and the correspondent 
inversion results (black circles) obtained using the same mode set 
as our observations given by Dynamics ML+HL.
The inversion results are very similar using any of the other two sets.
The two solar models are Model S and 
a model with identical physical assumptions but with a lower solar age, 4.52 Gyr, instead of 4.6 Gyr \citep{jcd96}.
The inversion results agree very well with the true model difference 
for $r \le 0.992$ $R_\odot$, specially taking 
into account the averaging kernel (dashed red line in the bottom panel).

\cite{howe96} showed that random noise in the observed mode frequency 
will introduce spurious oscillatory structures in the solution
as a result 
of the correlation of the errors in the solution at different radii.
The inversion result at two different radii makes use of similar data sets and their correspondent errors.
The solution errors are correlated even between points farther apart than the averaging-kernel width.
For examples of sound-speed error correlation,
see \cite{rabello99} and \cite{basu03b}.
To illustrate the effect of the error correlation in the inversion results presented here, it was 
carried out inversions of the theoretical frequency differences between the two models with noise added to them.
To each frequency difference the noise was generated as
normally distributed random numbers with the same standard deviation
as the error in the observed data.
The solid black line in Figure~\ref{artificial} shows the result for one realization.
The spurious oscillatory structures introduced by the correlated errors,
also present in Figure~\ref{fig:solution}, are clearly seen.
The two dotted lines correspond to the average of 1000 realizations plus and minus 2.5-$\sigma$,
showing that 
the overall results are consistent with the estimated formal errors. 

\begin{figure}[h]
\includegraphics
[scale=0.4]
{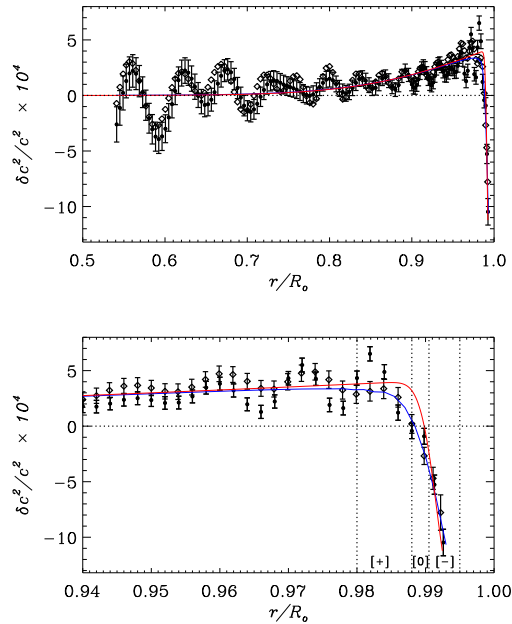}
\caption
{
Black circles and diamonds are the inversion results using Dynamics ML+HL and Dynamics ML respectively.
They are identical to the black and red circles in Figure~\ref{fig:solution}.
The red line is the smooth function, $\widetilde{\delta} c^2/c^2$, 
which was weighted by the averaging kernels and fitted to the inversion results of the two Dynamics sets.
The blue line is 
$\widetilde{\delta} c^2/c^2$
weighted by the averaging kernels.
}
\label{fig:gauss}
\end{figure}

To remove the spurious oscillation in our sound-speed determination,
a sum of two Gaussian functions was arbitrarily chosen as a smooth function of radius, $\widetilde{\delta} c^2/c^2(r)$.
It was `weighted' by the averaging kernels and fitted to the inversion results of the two Dynamics sets 
using a nonlinear least-square fit (red solid line in Figure~\ref{fig:gauss}).
According to 
this simple model,
the sound-speed difference increases with radius 
from around the base of the convection zone until 0.985 $R_\odot$ (i.e., 10 Mm below the surface)
at the second helium ionization zone
where $\widetilde{\delta} c^2/c^2 = 3.9 \times 10^{-4}$. 
Closer to the surface, it decreases steeply with radius.
It is zero at 0.990 $R_\odot$ (i.e., at a depth of 7 Mm) and, 
at 0.992 $R_\odot$ (or 5.5 Mm deep), $\widetilde{\delta} c^2/c^2$ = -9.2 $\times 10^{-4}$,
where the sound speed at solar maximum is smaller than at solar minimum.

\section{Discussion}

Using global helioseismology, \cite{baldner08} found an increase with 
radius in the sound-speed variation with solar cycle 
for 
$0.86 < r/R_\odot < 0.95$ (at the equator and at $15^\circ$ in latitude)
that agrees with our results within less than 1.5-$\sigma$.
Here, we found that the variation in sound speed increases until 
0.985 $R_\odot$ after which it decreases fast with radius becoming negative for $r > 0.990$ $R_\odot$. 
Posing the question of what is the physical mechanism behind this suddenly change in behavior at this particular depth.

This crossing of the sound speed at solar maximum 
from larger to smaller than at solar minimum 
is consistent in location with
the crossing of the sound speed below an active region from larger to smaller than at a quiet region obtained by ring-diagram analysis \citep{basu04, bogart08}.
The vertical lines in Figures~\ref{fig:solution} and~\ref{fig:gauss}
mark approximately the regions that correspond
to Figure~9 in \cite{bogart08} 
for positive, zero and negative sound-speed variations below an active region in relation to a quiet region.
This agreement 
indicates that
the sound-speed variation in the solar interior with solar cycle 
might be 
an overall effect of the active regions local perturbation, 
and not a global structure change with solar activity.
The estimated sound-speed variation 
obtained here is based on averages over
the whole Sun at a particular depth and over 
one or more solar rotations.

The number of active regions is very well correlated with the solar activity cycle.
Using the data provided by SIDC (Solar Influences Data Analysis Center of the Royal Observatory of Belgium),
the daily sunspot number averaged over the maximum-activity observing period (centered around May 2002) is about 120, 
while during the minimum (April 2008) is less than 10.
According to ring-diagram analysis,
the sound-speed variation at a depth of 14 Mm (i.e., 0.98 $R_\odot$) below 
a $16^\circ\times16^\circ$ region on the solar surface (tracked for 5.7 days) 
with a strong active region and with a magnetic activity index (MAI) of 
about 80 G in comparison to a quiet region is: 
$\delta c^2/c^2 \approx 0.01$ \citep{basu04}. 
For a smaller sunspot with an MAI of 40 G, it is 0.05.
In brief, the MAI is the average of MDI magnetograms over the 
same area in the solar surface and observing time as the analyzed region
\citep[for a full definition see][]{basu04}.
Performing a “back of the envelope” calculation,
if there are 11 regions with sizes of $16^\circ\times16^\circ$
(in the front and back of the Sun) with an MAI $\gtrsim 80$ 
during solar maximum (or a larger number of weaker regions), the mean sound-speed variation will be $\delta c^2/c^2 = 4 \times 10^{-4}$,
i.e., the value estimated here at this depth for the maximum-to-minimum solar-cycle variation.

Time distance is another local helioseismology technique that 
provides 
information about the subsurface structure of active regions.
While ring-diagram technique
is based on the analysis of three-dimensional power spectra 
in a small region of the Sun 
to measure the oscillation frequencies \citep{hill88}, 
as in global helioseismology,
time distance measures the wave travel times
between different points on the surface \citep{duvall97}.
Both the ring-diagram and time-distance techniques give qualitatively similar results, revealing a
characteristic two-layer seismic sound-speed structure.
However, the transition between the negative and positive variations occurs 
at different depths: $\sim$ 4 Mm for the time-distance result, 
and $\sim$ $5 - 8$ Mm for the ring-diagram inversions. 
It seems that this can be explained, at least in part,
by differences in the sensitivity and resolution of the two methods \citep[see][]{kosovichev11}.

Recently, \cite{Ilonidis} using time-distance technique detected strong acoustic travel-time anomalies around 65 Mm (i.e., at 0.91 $R_\odot$) below the solar surface where a sunspot emerged 1 to 2 days later.
These travel-time perturbations are an indication of sound-speed variations.
In our simple model, $\widetilde{\delta} c^2/c^2 = 2.1 \times 10^{-4}$ at this depth.
However, the observed signal lasts only for $\sim8$ hours
and is not as strong in the layers immediately above the detected signal.

In conclusion, using acoustic mode frequencies observed by MDI/SOHO through 
most of solar cycle 23, we have estimated the sound-speed difference between 
solar maximum and minimum from 0.53 to 0.992 $R_\odot$. 
The sound-speed difference seems to increase with radius from 
the base of the convection zone or above
until 0.985 $R_\odot$ (i.e., 10 Mm below the surface) at the second helium ionization zone.
Closer to the surface, it decreases steeply with radius.
It is zero at 0.990 $R_\odot$ (i.e., at a depth of 7 Mm) and, 
at 0.992 $R_\odot$ (or 5.5 Mm deep), 
$\widetilde{\delta} c^2/c^2 = -9.2 \times 10^{-4}$.
The transition of the sound speed at solar maximum from larger to smaller than at solar minimum
takes place at a similar depth as
the transition of the sound speed below an active region from larger to smaller than at a 
quiet region obtained by ring-diagram analysis, suggesting that
the sound-speed variation in the solar interior with solar cycle could be
an overall effect of the active regions local perturbation, 
and not a global structure change with solar activity.

\acknowledgements

We thank Jesper Schou and Tim Larson for providing
the spherical-harmonic decomposition of the MDI images and 
the frequency tables
from the MDI peak-fitting method.
SOHO is a project of international cooperation between ESA and NASA.



\end{document}